# Control of Intermolecular Bonds by Deposition Rates at Room Temperature: Hydrogen Bonds vs. Metal-Coordination in Trinitrile Monolayers

Thomas Sirtl,[†,o] Stefan Schlögl,[†,o] Atena Rastgoo-Lahrood,[†,o] Jelena Jelic,[◊] Subhadip Neogi,[§] Michael Schmittel,[§] Wolfgang M. Heckl,[o,†,‡] Karsten Reuter,[◊] and Markus Lackinger [o,‡,*]

[†] Department of Physics, Tech. Univ. Munich, James-Franck-Str. 1, 85748 Garching, Germany

[o] Center for NanoScience (CeNS), Schellingstr. 4, 80799 Munich, Germany

[◊] Department of Chemistry, Tech. Univ. Munich, Lichtenbergstr. 4, 85747 Garching, Germany

[§] Center of Micro- and Nanochemistry and Engineering, Organische Chemie I, Universität Siegen, Adolf-Reichwein-Str. 2, 57068 Siegen, Germany

[‡] Deutsches Museum, Museumsinsel 1, 80538 Munich, Germany

Supporting Information available

**ABSTRACT:** Self-assembled monolayers of 1,3,5-tris(4'-biphenyl-4''-carbonitrile)-benzene - a large functional trinitrile molecule - are studied on the (111) surfaces of copper and silver under ultra-high vacuum conditions by scanning tunneling microscopy (STM) and low energy electron diffraction (LEED). A densely packed hydrogen bonded polymorph was equally observed on both surfaces. Additionally, deposition onto Cu(111) yielded a well-ordered metal-coordinated porous polymorph that coexisted with the hydrogen bonded structure. The required coordination centers are supplied by the adatom gas of the Cu(111) surface. On Ag(111), however, the well-ordered metal-coordinated network was never observed. Differences in the adatom reactivity between copper and silver and the resulting bond strength of the respective coordination bond are held responsible for this substrate dependence. By utilizing ultra-low deposition rates, we demonstrate that on Cu(111) adatom kinetics plays a decisive role in the expression of intermolecular bonds – and hence for structure selection.

## INTRODUCTION

Self-assembled organic monolayers are promising candidates for the development of novel materials with tremendous options.[1] Many of their properties decisively depend on the type of intermolecular bond that stabilizes the monolayer.[2] The bond type is mostly predetermined by functionalization, but may additionally be influenced by kinetic effects. For instance, temperature can affect structure formation due to activation barriers.[3,4] Amongst the different non-covalent intermolecular bonds that can stabilize monolayers, metal-coordination is not only the strongest,[5] but, e.g. thiolate-copper coordination bonds can also offer strong intermolecular electronic coupling as required for molecular electronic applications.[6]

Formation of metal-coordination bonds requires both electron-donating ligands and metal centers. For surface-supported systems, the latter can be supplied either by metal deposition or by the adatom gas of a metal surface. Deposition of extrinsic metal centers facilitates chemical variability, while network formation with intrinsic metal centers offers facile preparation. Carboxylates and thiolates are suitable anionic ligands for coordinative bonds,[2,6-8] while nitrogen in heterocycles (e.g. pyridine or other azines) or nitriles are among the favored neutral ligands.[2,5,9,10] In contrast to thiol and carboxyl groups, where formation of metal-coordinated networks requires additional thermal activation,[2,6,8] nitrile coordination is readily observed at room temperature with an onset around 180 K.[3]

For many correspondingly functionalized molecules the type of intermolecular bond can be changed by supplying coordination-centers, as e.g. for dinitriles on Ag(111).[5,9,11-13] Additional deposition of cobalt atoms induces a change from hydrogen bonding to metal-coordination, accompanied by structural reorganization.

Besides temperature, the competition between molecular flux onto the surface and diffusion on the surface can also play a decisive kinetic role in determining the structure. An experimental example is given by Li et al., who observed structurally different pyridyl-porphyrin monolayers upon variation of the deposition rate.[14] Yet, all of these structures were stabilized by the same type of intermolecular bond, i.e. copper coordination and no change of bond type was induced.

Here we study self-assembly of the large functional molecule 1,3,5-tris(4'-biphenyl-4''-carbonitrile)-benzene (BCNB) on both Cu(111) and Ag(111). These surfaces were chosen as substrates because both exhibit 2D adatom gases that are comparable in mobility,[15] but differ in reactivity.[4,16] In order to study the influence of the above mentioned kinetic competition on formation of adatom-coordinated trinitrile networks, experiments were conducted with a variation in deposition rate over two orders of magnitude.

## EXPERIMENTAL DETAILS

BCNB monolayers were characterized by scanning tunneling microscopy (STM) and low energy electron diffraction (LEED)

in ultra-high vacuum (UHV). STM data were acquired with a home-built beetle type STM driven by a SPM100 controller from RHK. The topographs were processed by a mean value filter. All images were obtained at room temperature at a base pressure below $3\times10^{-10}$ mbar. Both Ag(111) and Cu(111) single crystal surfaces were prepared by cycles of Ne$^+$ ion-sputtering at 1 keV and e-beam annealing at 550 °C for 30 minutes. Thorough calibration of the STM with atomically resolved topographs of Cu(111) allows to derive lattice parameters and distances with an accuracy of ~5%.

LEED experiments were performed in a separate UHV chamber at a base pressure below $1\times10^{-10}$ mbar. The LEED optics (Omicron NanoTechnology GmbH) was controlled by an electronics from SPECS Surface Nano Analysis GmbH. Ag(111) and Cu(111) surfaces were prepared by Ar$^+$ ion sputtering at 2 keV and subsequent e-beam annealing at 550 °C for 30 min. Deposition parameters were similar to STM experiments. LEED patterns were acquired at a sample temperature of ~60 K. The software LEEDpat3 was used for geometric simulations.

1,3,5-tris(4'-biphenyl-4''-carbonitrile)-benzene (BCNB, cf. inset in Figure 1 for molecular structure and Supporting Information for synthesis) was deposited from a home-built Knudsen-cell[17] and thoroughly outgassed prior to deposition. The substrates were held at room temperature. The crucible temperature was varied between 280 °C and 330 °C, resulting in deposition rates between ~$2.5\times10^{-4}$ and ~$2.5\times10^{-2}$ monolayer/min, respectively. In order to determine the BCNB sublimation rate vs. crucible temperature and to verify the long-term stability of the sublimation, a Knudsen-cell equipped with a quartz crystal microbalance was used.[18]

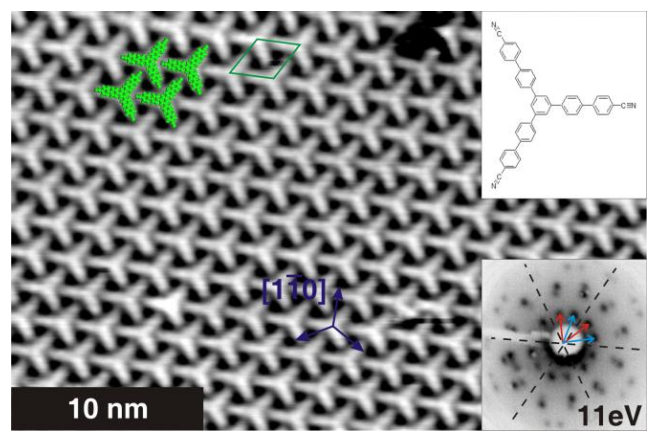

Figure 1. STM topograph of the densely packed ($\sqrt{39}\times\sqrt{39}$)R±16° BCNB superstructure on Ag(111) (+1.61 V, 100 pA). Molecules are overlaid and the unit cell is indicated by green lines. The two rotational domains are marked in the LEED pattern (lower inset) by red and turquoise arrows, respectively.[18] Dashed lines indicate high symmetry substrate directions.

## RESULTS AND DISCUSSION

On Ag(111) BCNB self-assembles into long-range ordered, densely packed monolayers with $p31m$ symmetry. Both STM and LEED consistently reveal a ($\sqrt{39}\times\sqrt{39}$)R±16° superstructure with a lattice parameter of 1.80 nm (cf. Figure 1). The existence of two rotational domains is evident in the LEED pattern depicted in the lower right inset.[18] Identification of the molecular arrangement is unambiguous, since the threefold contour of BCNB is clearly recognizable and the STM-derived size is in excellent agreement with the optimized geometry of isolated molecules.[18] Given the dense packing of BCNB, additional constituents can be excluded. Based on the molecular arrangement, it is concluded that the dominant interactions are C≡N···H−C hydrogen bonds as similarly observed in surface self-assembly[3,11,19] and bulk crystals[20,21] of carbonitriles. The nitrile groups are in close proximity to three phenyl hydrogen atoms. DFT calculations of two isolated molecules result in a center-to-center distance of 1.81 nm.[18] The extremely small deviation from the experiment (1.80 nm) indicates a minor substrate influence and justifies comparison with gas phase calculations. The closest N···H distances (260 pm) are consistent with hydrogen bond lengths of comparable bulk crystals (250 pm - 260 pm).[20,21]

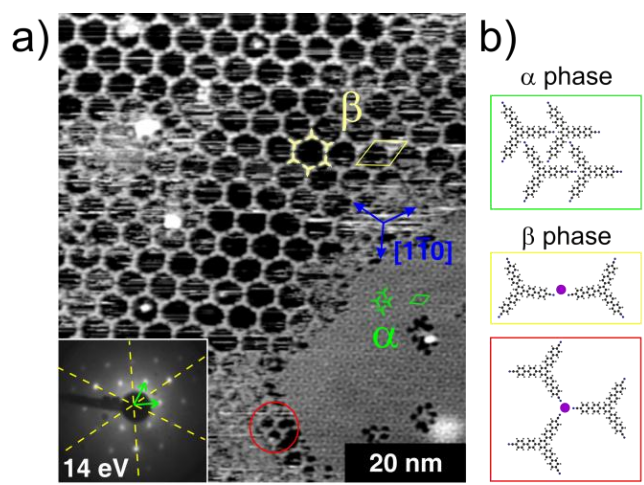

Figure 2. a) STM topograph of a BCNB monolayer on Cu(111) (+2.09 V, 80 pA) deposited with a rate of ~$2.5\times10^{-2}$ monolayer/min. Inset: LEED pattern,[18] arrows indicate the reciprocal unit cell vectors of the α phase, dashed lines mark high symmetry substrate directions. b) Tentative binding models of α (green) and β (yellow) phase; occasionally threefold coordination (red) was also observed, but only in isolated arrangements.

BCNB deposition with a rate of ~$2.5\times10^{-2}$ monolayer/min onto Cu(111) yielded two polymorphs, both with $p31m$ symmetry. The overview STM image in Figure 2a illustrates the coexistence of both a densely packed α and a porous β phase. Lattice parameter and molecular arrangement of the α phase on Cu(111) are comparable to those on Ag(111). The commensurate ($4\sqrt{3}\times4\sqrt{3}$)R30° superstructure on Cu(111) (cf. inset in Figure 2a for a LEED pattern)[18] exhibits a slightly smaller lattice parameter of 1.77 nm and only one rotational domain. According to the similarities with the structure on Ag(111) it is concluded that the α phase on Cu(111) is likewise stabilized by similar intermolecular C≡N···H−C hydrogen bonds.

The pores of the β phase are arranged on a hexagonal lattice and the unit cell contains two molecules. The streaky features that are observed within some, but not all pores arise from entrapped mobile species, either excess molecules or adatoms. Similar observations were reported for various porous systems.[11,22-25] Two different epitaxial relations to the substrate – a (11√3×11√3)R±30° and a 19×19 superstructure – were found, that are almost identical in lattice parameters (4.87 nm vs. 4.86 nm, respectively). Generally, emergence of a porous polymorph already hints towards stronger intermolecular bonds. From the rather large center-to-center distance of adjacent molecules of 2.8 nm direct interactions via intermolecular hydrogen bonds can a priori be excluded. A detailed view on the β phase furthermore reveals an arrangement where the molecular lobes are not aligned with the long diagonal of the unit cell, but slightly tilted by ±9° (cf. Figure 3). In most dimeric binding motifs, the two molecules tilt into the same direction, i.e. one molecule tilts clockwise, the other counter-clockwise. Tilts can occur in both directions and occur in segregated domains.[18]

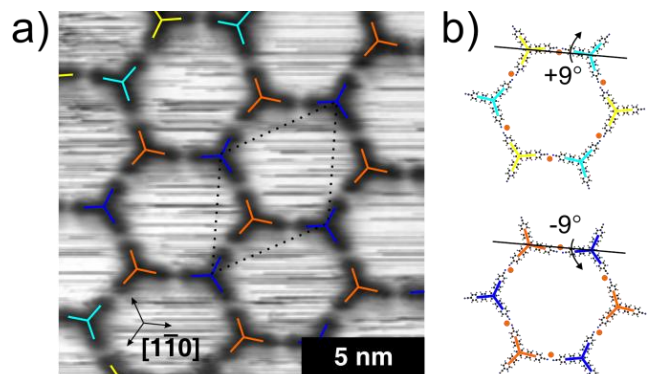

Figure 3. β phase of BCNB on Cu(111). a) Close-up STM topograph (1.41 V, 80 pA) with overlaid tripods, the color encodes the different tilts by ±9°. b) Tentative models. Carbon: grey, copper: orange, hydrogen: white, nitrogen: blue.

Based on the intermolecular arrangement we propose that BCNB molecules are interconnected by coordination bonds of the nitrile groups with copper adatoms. This hypothesis is further substantiated by occasionally observed adatom related contrast features in STM images, DFT calculations, and the good match with an epitaxial model, as detailed in the following. In accord with most other experiments on copper-coordinated networks, the copper atoms are normally not resolved by STM.[26-32] This invisibility of obviously present coordination centers in STM images is also reported for many other metal-coordinated networks, such as Co-coordinated nitriles[5,9,12,33] and attributed to an electronic effect.[34] Nevertheless, occasionally for peculiar tip conditions, distinct topographic maxima were observed in the β phase at the proposed position of the copper coordination centers, i.e. midway between two BCNB molecules (cf. Supporting Information). Moreover, in these images, the BNCB molecules appear with diminished apparent height, indicating a tip that is sensitive to electronic states in a different energy range. Owing to their position these topographic maxima are unambiguously identified as the coordinating copper adatoms. Similar signatures of coordinating adatoms in the STM contrast have been reported for Cu-benzoate complexes on Cu(110).[35]

In the β phase, copper adatoms coordinate two nitrile groups. Threefold coordination, the major binding motif in Co-coordinated nitrile networks[5,9,12] and hitherto known Cu-coordination,[3] is instead only rarely observed in isolated arrangements (cf. Figure 2). However, for surface-supported metal-coordinated networks, unusual coordination with lower coordination numbers seems to be the more general case.[8,27,36] This can be rationalized by the special environment of these surface-supported systems. On the one hand there is the restriction to a planar geometry due to the surface confinement, on the other hand there is an additional electronic influence of the metal surface due to charge transfer and screening by the free electrons.[27,36]

DFT calculations were performed, in order to derive optimal bond lengths for copper-nitrile coordination and to find explanations for the tilt in intermolecular bonds. For a full account of surface effects, it is also important to include the copper substrate. However, due to the large system size, the DFT calculations had to be restricted to benzonitrile as a representative model system. As depicted in Figure 4, energy minimization of two benzonitrile molecules coordinated to one copper adatom, results in a straight bond with the adatom stably adsorbed in a three-fold hollow site. The N-Cu bond length amounts to 0.192 nm, and the copper adatom resides 0.200 nm above the topmost copper layer and 0.121 nm below the benzene rings. The DFT calculations reveal a global energetic minimum for the benzene rings oriented along the <1-10> high symmetry direction of the substrate, and a further, only slightly less stable local minimum for alignment in the bisecting <11-2> directions.

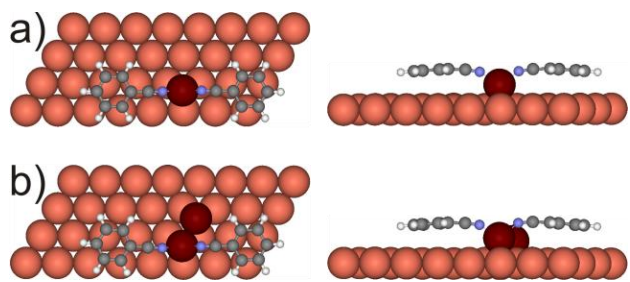

Figure 4. DFT geometry optimization of two benzonitrile molecules coordinated to copper adatoms (represented by dark red spheres). Left: top-view, i.e. parallel to [111]; right: side-view, i.e. parallel to [11-2]; only one copper substrate layer is depicted, but three layers were considered in the calculation. a) Coordination by one copper adatom in a three-fold hollow site. b) Coordination by one copper adatom in a three-fold hollow site with a further copper adatom in an adjacent site.

Since DFT geometry optimization in either case yields a straight bond, the experimentally observed tilt cannot be explained by intrinsic properties of the chosen model system. In an alternative approach, a second copper adatom was placed adjacent to the coordinating copper atom, likewise into a three-fold hollow site. The optimized geometry is depicted in Figure 4b. Addition of a second adatom actually results in a tilt of the bond angle by 6°, thereby offering a possible explanation.

Although the DFT simulations of the model systems include direct substrate effects, conceivable registry effects arise for the full BCNB molecule that can be relevant for the β phase. Nevertheless, from the DFT calculations an optimal N-Cu bond length in the adsorbed system of 0.192 nm was deduced and there is a clear confirmation that copper adatoms have a strong preference for three-fold hollow sites. The fairly large unit cell of the β phase contains two BCNB molecules and three copper adatoms. When the above stated requirements are considered, it becomes clear that the substrate registry does not allow for a straight bond configuration. In order to keep C≡N-Cu coordination bonds within 5 % of the optimized length and guarantee three-fold hollow sites for all adatoms, the β phase has to adapt to the substrate lattice in the energetically most efficient way, i.e. by tilting the BCNB molecules with respect to each other. A tentative model of the β phase that takes all requirements and the experimental tilt angle into account is presented in Figure 5.

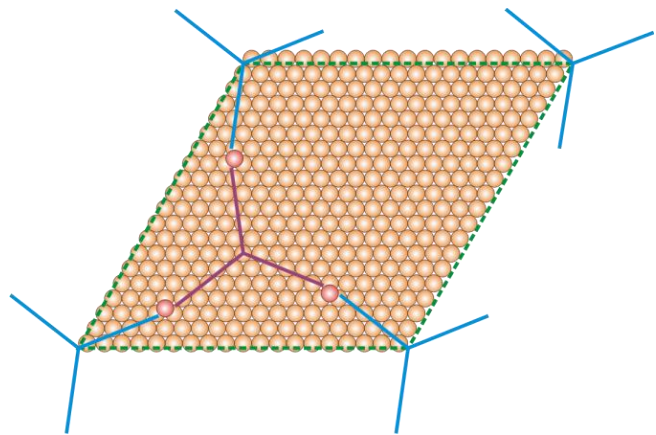

Figure 5. Tentative model of the β phase on Cu(111), the dashed green lines indicate the 19 × 19 unit cell. For clarity the BCNB molecules are represented by tripods. This model considers the experimental tilt angle and the preference of copper adatoms for three-fold hollow sites. All C≡N-Cu are within 5% of the DFT derived optimal bond length of 0.192 nm. The three coordinating copper adatoms are located on the same sublattice.[18]

In order to obtain insights into the growth kinetics, experiments were conducted with ultra-slow deposition, i.e. a surface coverage of one monolayer is accomplished in 67 h. On Cu(111) this results in exclusive formation of the β phase, thereby hinting to a kinetic origin of the polymorphism. A representative STM image is presented in Figure 6a. Irrespective of the slow deposition rate, a similar tilt angle was observed between the BCNB molecules, pointing to an equilibrium effect. Occasionally, the STM images also show a parallel side-by-side arrangement of two dimers, an example is highlighted in Figure 6a. Yet, these uncommon coordination schemes occur grouped along a line, and are thus attributed to an antiphase domain boundary.

Interestingly, a similarly slow deposition onto Ag(111) still resulted in the densely packed polymorph, yet with notably extended domain size, as illustrated in Figure 6b.

We thus assign the polymorphism on Cu(111) to a kinetic effect, namely the availability of Cu adatoms. Upon deposition of BCNB, formation of the β phase consumes Cu adatoms and thus depletes the density of the adatom gas below its equilibrium value. If the progressive consumption of Cu adatoms caused by further deposition of BCNB molecules is faster than adatom replenishment from step-edges, the availability of Cu centers for coordination bonds decreases. In the absence of Cu adatoms the second best option in terms of intermolecular bonds is realized, i.e. the hydrogen bonded α phase.

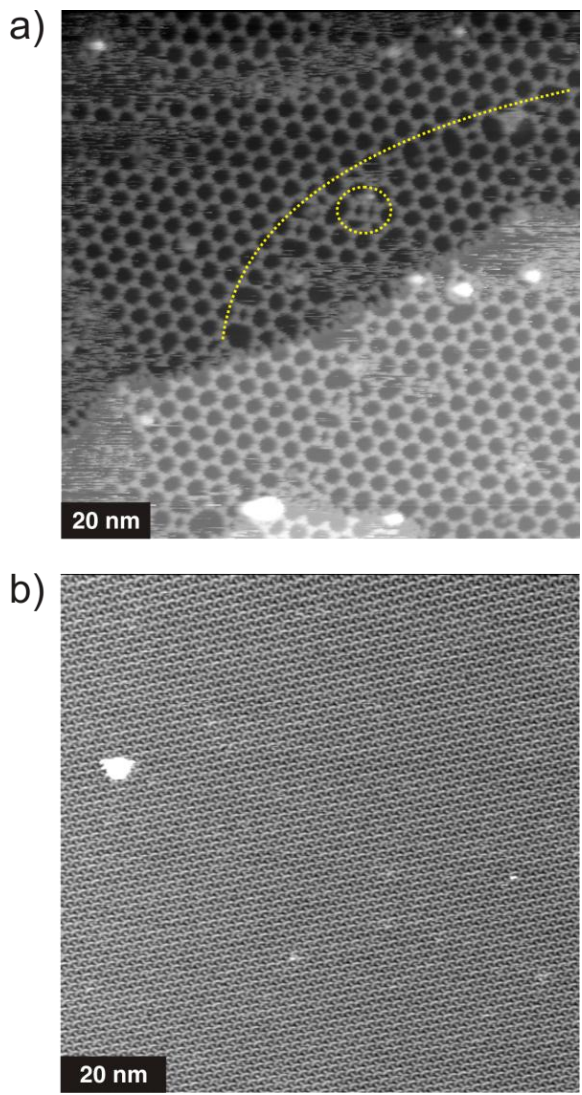

Figure 6. a) STM topograph of BCNB on Cu(111) (+2.01 V, 39 pA). The monolayer was prepared by ultra-slow deposition (~2.5×10$^{-4}$ monolayer/min), yielding exclusively the porous β phase. The dashed circle highlights a parallel side-by-side arrangement of dimers, the dashed line indicates a domain boundary b) STM topograph of BCNB on Ag(111) (−0.19 V, 40 pA). The monolayer was similarly prepared by ultra-slow deposition (~8.3×10$^{-4}$ monolayer/min). However, on Ag(111) no change of intermolecular bond type was induced. Exclusively the α phase is observed, yet with notably increased domain size.

This picture is supported by the experiments on Cu(111) with ultra-slow deposition, resulting in the exclusive formation of the β phase. When the BCNB deposition rate is so low that the equilibrium density of the adatom gas is not perturbed, Cu centers are constantly available and preferred formation of Cu-coordination bonds is not hampered by kinetic limitations. In contrast, the insufficient reactivity of Ag adatoms at room temperature leads to the exclusive formation of the hydrogen bonded structure on Ag(111), even for ultra-slow deposition. On this less reactive surface, the slow deposition only affected the nu-cleation and growth kinetics, resulting in extended domains of the densely packed hydrogen bonded phase.

This striking difference between Cu(111) and Ag(111) can be explained by the lower bond dissociation energy (BDE) of C≡N-Ag vs. C≡N-Cu coordination bonds. The BDE of two benzonitrile molecules coordinated either by one Cu or Ag atom was evaluated for isolated arrangements by DFT calculations. The BDE for the copper case amounts to 0.90 eV per benzonitrile molecule, and is substantially higher than 0.30 eV for the silver case. Accordingly, Ag coordinated BCNB networks might only be stable at lower temperature, however, then the adatom density and mobility become the limiting factors.

## CONCLUSION

The deposition-rate dependent study of BCNB self-assembly on Ag(111) and Cu(111) demonstrates the importance of both substrate and kinetic effects for the expression of a specific type of intermolecular bond. Two types of intermolecular bonds dominate in BCNB monolayers, namely hydrogen bonds and metal-coordination bonds with adatoms. Their emergence can be controlled by the choice of substrate, a well-known effect that is in the case of metal-coordination mostly related to adatom reactivity and the bond strength of the respective coordination bonds. Moreover, here we discover that also the deposition rate is effective in deliberately selecting the type of intermolecular bond, and thus, controlling the structure. A qualitative study of these kinetic effects allows for a basic understanding of growth kinetics and polymorph selection in abundantly employed formation of metal-organic networks through adatom coordination.

## ASSOCIATED CONTENT

**Supporting Information.** Synthesis details, sublimation rate vs. temperature and time curves, additional DFT calculations, STM and LEED data, and an additional structure model. This material is available free of charge via the Internet at http://pubs.acs.org.

## AUTHOR INFORMATION

**Corresponding Author**
* markus@lackinger.org, http://www.2d-materials.com

## ACKNOWLEDGMENT

Funding by the Fonds der Chemischen Industrie (T. S.), the Nanosystems-Initiative-Munich cluster of excellence, Alexander von Humboldt Foundation (S. N.) and Elitenetzwerk Bayern (St. S.) and discussions with Debabrata Samanta are gratefully acknowledged.

Insert Table of Contents artwork here

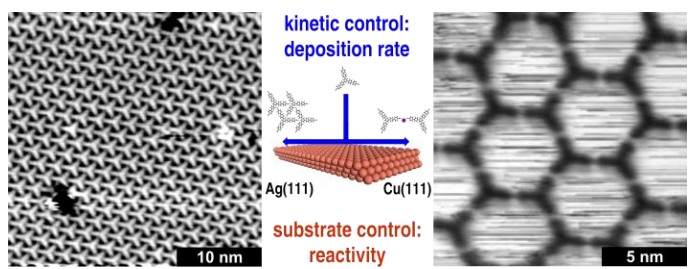